\documentclass[aoas,preprint]{imsart}
\setattribute{journal}{name}{}
\RequirePackage[OT1]{fontenc}
\RequirePackage{amsthm,amsmath,natbib,calligra}
\RequirePackage[colorlinks,linkcolor=blue,citecolor=blue,urlcolor=blue]{hyperref}
\RequirePackage{natbib}
\RequirePackage{comment}
\RequirePackage{graphicx,subfigure,latexsym,amssymb}
\RequirePackage{float,epsfig,multirow,rotating,times}
\RequirePackage{upgreek,wrapfig}

\arxiv{stat.AP}
\startlocaldefs

\numberwithin{equation}{section}

\newcommand{\bY}{\boldsymbol{Y}}
\newcommand{\by}{\boldsymbol{y}}
\newcommand{\bX}{\boldsymbol{X}}
\newcommand{\bI}{\boldsymbol{I}}
\newcommand{\bx}{\boldsymbol{x}}
\newcommand{\bp}{\boldsymbol{p}}
\newcommand{\bV}{\boldsymbol{V}}
\begin{document}

\begin{frontmatter}
\title{Minimum Distance Estimation of Milky Way Model Parameters and Related Inference}
\runtitle{Minimum Distance Estimation of Milky Way Model Parameters}

\begin{aug}
\author{
{\fnms{Sourabh} \snm{Banerjee}\thanksref{t1}\ead[label=e1]{sbanerj7@illinois.edu}},
{\fnms{Ayanendranath} \snm{Basu}\thanksref{t2}\ead[label=e2]{ayanbasu@isical.ac.in}},
{\fnms{Sourabh} \snm{Bhattacharya}\thanksref{t3}\ead[label=e3]{sourabh@isical.ac.in}},
{\fnms{Smarajit} \snm{Bose}\thanksref{t2}\ead[label=e4]{smarajit@isical.ac.in}},
{\fnms{Dalia} \snm{Chakrabarty}\thanksref{t4,t5}\ead[label=e5]{d.chakrabarty@warwick.ac.uk}\ead[label=e6]{dc252@le.ac.uk}} and
{\fnms{Soumendu Sundar} \snm{Mukherjee}\thanksref{t6}\ead[label=e7]{soumendu@berkeley.edu}}
}

\thankstext{t1}{Graduate Student at Department of Statistics, University of Illinois at Urbana-Champaign}
\thankstext{t2}{Professor at Bayesian and Interdisciplinary Research Unit, Indian Statistical Institute}
\thankstext{t3}{Assistant Professor at Bayesian and Interdisciplinary Research Unit, Indian Statistical Institute}
\thankstext{t4}{Associate Research fellow at Department of Statistics,
  University of Warwick}
\thankstext{t5}{Lecturer of Statistics at Department of Mathematics,
  University of Leicester}
\thankstext{t6}{Graduate Student at Department of Statistics, University of California, Berkeley}

\runauthor{Banerjee, Basu, Bhattacharya, Bose, Chakrabarty and Mukherjee}

\address{
Department of Statistics\\
University of Illinois at Urbana-Champaign\\
725 South Wright Street\\
Champaign, Illinois 61820, USA\\
\printead*{e1}\\
\and\\
Bayesian and Interdisciplinary Research Unit\\ 
Indian Statistical Institute\\
203, B. T. Road\\ 
Kolkata 700108, India\\
\printead*{e2}\\
\printead*{e3}\\
\printead*{e4}\\
\and\\
Department of Statistics\\
University of Warwick\\
Coventry CV4 7AL,
U.K.\\
\printead*{e5}\\
\and\\
Department of Mathematics\\
University of Leicester \\
Leicester LE1 7RH,
U.K.\\
\printead*{e6}\\
\and\\
Department of Statistics\\
University of California, Berkeley\\
367 Evans Hall\\
Berkeley, CA 94720-3860, USA\\
\printead*{e7}\\
}
\end{aug}

\begin{abstract}
We propose a method to estimate the location of the Sun in the disk of
the Milky Way using a method based on the Hellinger distance and
construct confidence sets on our estimate of the unknown location
using a bootstrap based method. Assuming the Galactic disk to be
two-dimensional, the sought solar location then reduces to the radial
distance separating the Sun from the Galactic center and the angular
separation of the Galactic center to Sun line, from a pre-fixed line on
the disk. On astronomical scales, the unknown solar location is
equivalent to the location of us earthlings who observe the velocities
of a sample of stars in the neighborhood of the Sun. This unknown
location is estimated by undertaking pairwise comparisons of the
estimated density of the observed set of velocities of the sampled
stars, with the density estimated using synthetic stellar velocity
data sets generated at chosen locations in the Milky Way disk. The
synthetic data sets are generated at a number of locations that we
choose from within a constructed grid, at four different base
astrophysical models of the Galaxy. Thus, we work with one observed
stellar velocity data and four distinct sets of simulated data
comprising a number of synthetic velocity data vectors, each generated
at a chosen location. For a given base astrophysical model that gives
rise to one such simulated data set, the chosen location within our
constructed grid at which the estimated density of the generated
synthetic data best matches the density of the observed data, is used
as an estimate for the location at which the observed data was
realized. In other words, the chosen location corresponding to the
highest match offers an estimate of the solar coordinates in the Milky
Way disk. The ``match'' between the pair of estimated densities is
parameterized by the affinity measure based on the familiar Hellinger
distance. We perform a novel cross-validation procedure to establish a
desirable ``consistency" property of the proposed method.
\end{abstract}



\end{frontmatter}

\section{Introduction and Background}\label{SEC:intro}

\paragraph{}
The learning of structure in the space of parameters of a system,
using available data, is an exercise that has gained increasing
attention in the recent past. This includes attempts at finding
inter-variable relationships in large data using developed methods of
scoring the association \citep{reshef}, in graphical model contexts
\citep{heckerman_95, zoubin}, by searching for chosen features
within the data \citep{lee,zomorodian_04}, by developing
high-dimensional regression models in regression frameworks
characterized by the number of covariates far exceeding the number of
responses \citep{yuanlin_07, simon_12}, and by developing density based
distances within the paradigm of semi-supervised or unsupervised learning 
\citep{bijral_2011,sajama_05,weinberg_04}.

\paragraph{}
Indeed, unsupervised learning is often the relevant framework in
real-world problems. However, within the framework of supervised
learning, the aim is to predict values of the response variable $\bY$
corresponding to a given set of predictor variables $\bX$, given the
training sample $(\bx_1, \by_1^{(\star)})$, $(\bx_2, \by_2^{(\star)})$,
$\ldots,$ $(\bx_n, \by_n^{(\star)})$. Here $\by_i^{(\star)}$ is a known
value of $\bY$ at a chosen value $\bx_i$ of $\bX$. Here, the $i$-th such 
chosen $\bx_i$ is the $i$-th design vector. The aim is to learn
the model parameters that minimize the expected loss at each
$\bx$, where the loss function is appropriately chosen to embody the
error in the estimation of the value of the response variable
\citep{hastie_book}.  The probability density of the response variable
$\bY$, conditional on $\bX$ is considered with the aim of learning the
unknown model parameters. In contrast, in the framework of
unsupervised learning, the joint probability density of the
observations is examined, with the aim of making inference on the
model parameters.

\paragraph{}
In this paper, we present a novel application in which the aim is to
perform estimation of the unknown model parameters by comparing the
conditional density of synthetic values of the response variable given
a chosen set of predictor values with that of the measured values of
the response variable given the same predictor set. Though this
resonates with the supervised learning scheme, there are some features
of this implementation that mark it as atypical in terms of a
supervised learning scheme. Firstly, here the response variable $\bY$
is a matrix; it is more the case in unsupervised learning that $\bY$
is a high-dimensional variable. Secondly, in this work, the loss
function is itself defined in terms of the distance between the two
aforementioned conditional probability density functions; the $\bx$
for which this distance is minimized, gives the unknown model
parameter. Thirdly, the density functions in question are not known to
begin with but are estimated using kernel density estimation
techniques. In fact, the estimated densities are found to be highly multimodal as well as sparse. The efficiency of this learning may, however, be compromised if the chosen minimum distance procedure is not robust against violations of the usual model assumptions \citep{basu}.

\paragraph{}
In particular, we invoke an affinity measure based on the Hellinger
distance, between the densities that the observed data and the
synthetic data are sampled from. The motivating idea in this work is
that the synthetic data sets are realizations of simulations of the
system under a variety of given values of the model parameter
vector. Thus, the particular synthetic data set that maximizes the
affinity between the said densities is the realization obtained from
the model parameter value that corresponds best to the true value;
the ``true'' value of the model parameter indicates the value which
suitably describes the observations. Maximization of the affinity in 
this context is equivalent to the minimization of the Hellinger distance.  

\paragraph{}
One fundamentally important aspect of statistical learning is to
perform model selection \citep{kohavi95, kearns97} and importantly to
quantify accuracy of a given model, using available data
\citep{marklast}.  It is in principle possible to extend parameter
estimation using minimized Hellinger distance to higher dimensions
\citep{tamura86}. The accompanying parameter uncertainty estimation is
possible by constructing a high dimensional confidence set within the
region of interest, as distinguished from a product of confidence
intervals of interest along each dimension. In our application we seek
similarly constructed confidence set on our estimate of the unknown
parameters using a bootstrap based method.  It is also of vital
importance to ensure generalization of the learnt model to an
independent data set and is achieved using cross-validation techniques
\citep{efron}.  We include such validation of our learnt model
parameters by adopting a cross-validation technique where assuming a
particular location as the true location we verify whether they are
accurately estimated by the proposed method.

\paragraph{}
The paper is organized as follows. In Section \ref{SEC:exp_setup} we
describe the experimental set up under which the data are
generated. Some discussion of the existing literature related to this
problem is presented in Section \ref{SEC:literature}. The method we
advocate is described in Section \ref{SEC:proposed_method}. Section
\ref{SEC:results} contains the results of our analysis. In particular,
Section~\ref{SEC:confidence} discusses the bootstrap based method that
we use to construct confidence set on our estimation of the Milky Way
parameters and in Section~\ref{SEC:validation} we present our
implementation of cross-validation. Finally, Section
\ref{SEC:conclusion} provides some concluding remarks.

\section{The Experimental Set Up}\label{SEC:exp_setup}

\paragraph{}
In this application, the system under consideration is the disk of the
Milky Way that is assumed to be two-dimensional. The observed data
comprise the $N\times 2$-dimensional matrix $\bY =
(\by_1:\by_2:\ldots:\by_N)^T$, where $\by_j$ is a
two-dimensional velocity vector, $j=1,2,\ldots,N$. Thus $\bY$
represents the two-component velocity vector measurement of $N$ stars
that were observed close to the Sun in our galaxy \citep{fux}.  
For this astronomical observational data set, we have $N$=3500. 

\paragraph{}
Such a matrix of these velocity measurements is realized at location
$\bX$ of the observer who measures the velocities of these $N$
stars. Non-linear dynamical simulations of the Milky Way disk was
performed by \citet{chak1}, by varying this physical location
$\bX$. We place the two-dimensional Milky Way disk on a 2-dimensional
polar coordinate system such that the spatial location vector $\bX$ is
given by the radial distance $R$ from the defined center of this
coordinate system (chosen to coincide with the center of the Milky Way
disk) and the azimuthal or angular displacement $\theta$ (where
$\theta$=0 is chosen to be along the long axis of a feature in the
Milky Way, namely the central bar in the Galaxy). Thus, the value of
$\bX$ in a 2-dimensional orthogonal basis is
$\bx=(r\cos\theta,r\sin\theta)^T$. In fact, in our work, it is this
physical location $\bX$ of the observer on the two-dimensional Milky
Way disk that we want to learn. Thus, in this set up, what we referred
to as our ``unknown model parameters'' in the introductory section,
concurs with the unknown physical location of the observer. We would
like to emphasize that hereafter, ``location'' would refer to the
address of the observer on the Milky Way disk parameterized by
${\bX}$. According to our model, the observed velocity matrix
$\bY^{(obsvd)}$ corresponds to an unknown value of the location, i.e. at
${\bX}=\bx_\star=(r_\star\cos\theta_\star,r_\star\sin\theta_\star)^T$.

\paragraph{}
Thus, the identification of $\bx_\star$ is equivalent to identifying
the radial location $r_\star$ of the observer from the center of the
Galaxy and the angular location (separation) $\theta_\star$ of the
observer from a chosen axis in the Galaxy, such that if from this
location in the model Milky Way, the observer had tracked the stars in
the neighborhood of the Sun for their velocity vectors, the collected
data would have been ``closest'' to the observed data $\bY^{(obsvd)}$;
here the aforementioned ``closeness'' is in the sense implied by our
affinity measure (see Section \ref{SEC:distance_measure}). Now, the
location of the observer, is really {\it our} location as earthlings
on the Galactic disk, i.e. seeking
$(r_\star\cos\theta_\star,r_\star\sin\theta_\star)^T$ is the same as
trying to estimate the location of the Sun in the Milky Way
disk\footnotemark.  \footnotetext{On galactic length scales, the
  location of us, i.e. the Earth in the Galaxy, is very well
  approximated by the location of the Sun in the Galaxy.}

\begin{figure}[t!]
\centering{
  $\begin{array}{c c}                                                         
   \includegraphics[scale=0.35]{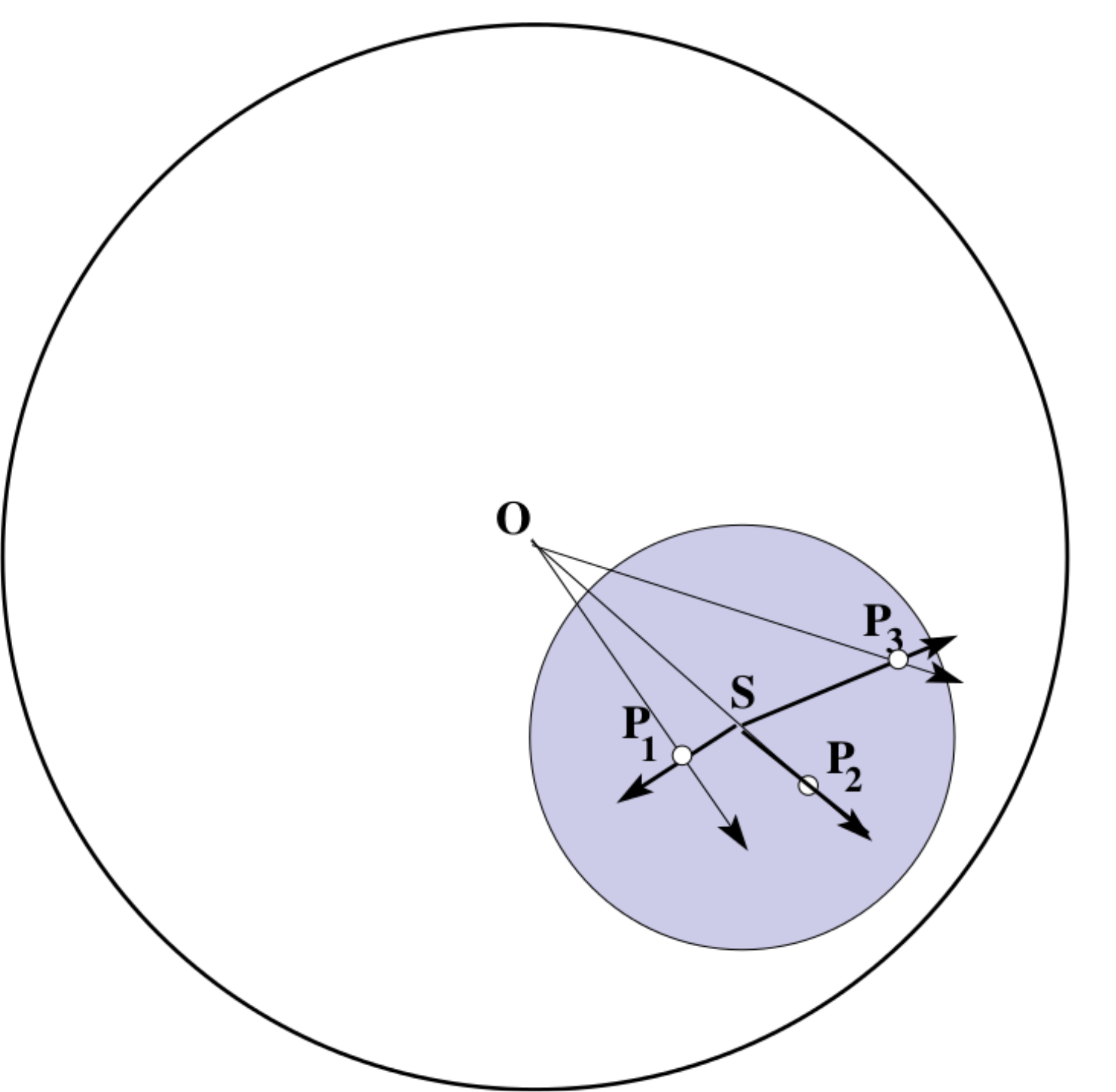} &                            
   \includegraphics[scale=0.35]{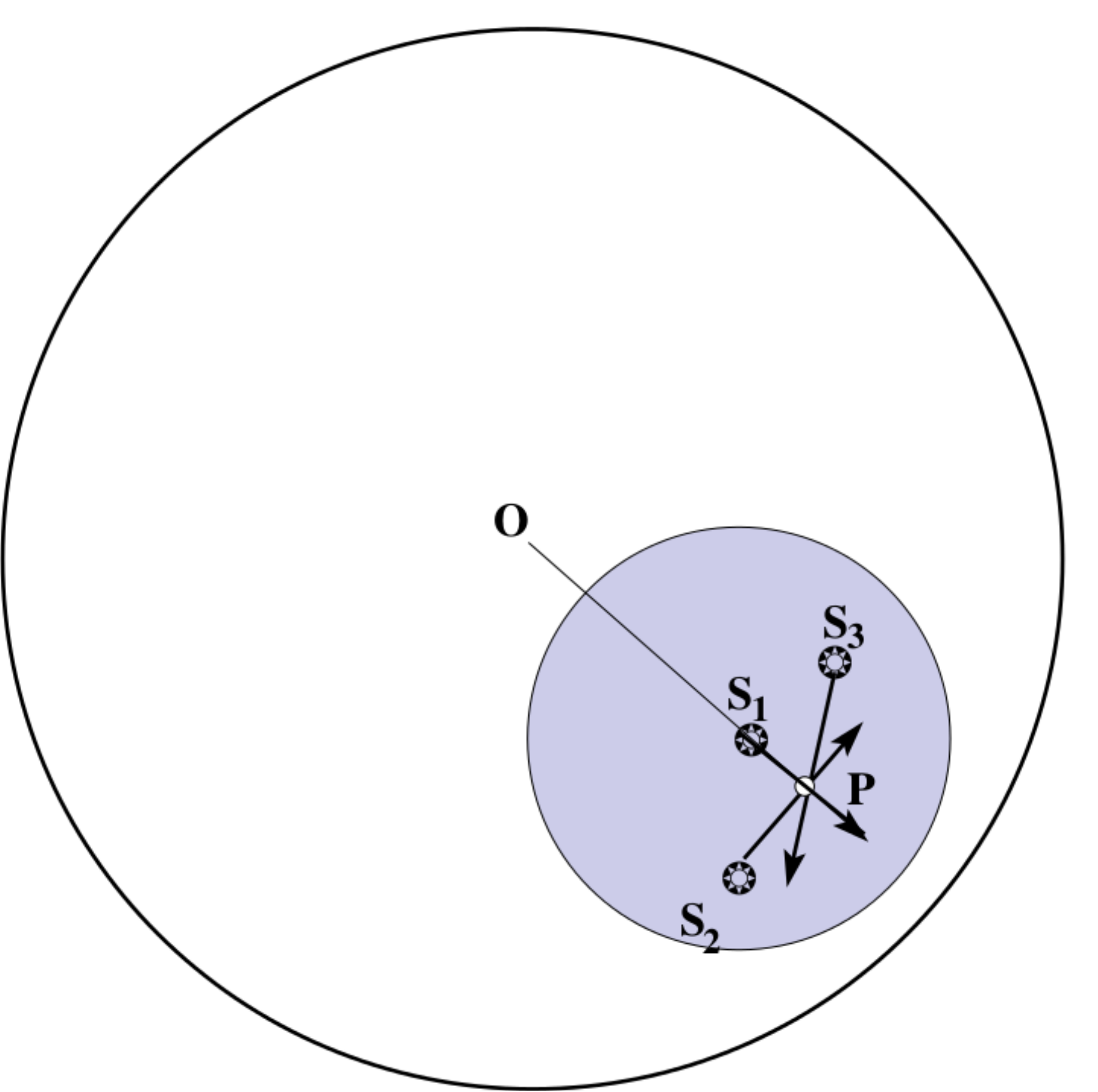}    
   \end{array}$
}
\caption{{\it Left:} figure showing locations of 3 of the
  sampled stars from center of circular patch (grey circle with center
  at location of the Sun--depicted at point ${\bf S}$) within which
  stars are sampled uniformly from. The location and velocity vectors of
  these sampled stars are recorded by an observer at the Sun. However,
  the locations of these sampled stars with respect to the center of
  the Galaxy (at ${\bf O}$) are unknown. Thus, a measured
  heliocentric stellar location vector cannot constrain the unknown
  location vector of the Sun with respect to the center of the Galaxy
  (the vector ${\bf{OS}}$). Neither can the distribution of the
  measured heliocentric stellar locations constrain the unknown
  ${\bf{OS}}$ since the sample mean of the measured heliocentric stellar
  locations is zero. {\it Right:} velocity $\bV$ of an example star (at ${\bf
    P}$) along ${\bf{OP}}$ is viewed by observer at point ${\bf S_1}$
  to lie entirely along the line that joins this observer to the star
  while the observer at point ${\bf S_2}$ views this stellar velocity
  to be entirely orthogonal to her line-of-sight to the star at ${\bf
    P}$. The velocity vector of a sampled star, as measured by
  an observer, is expressed to comprise a radial component that is
  along the line-of-sight of the observer to the star, and a
  transverse component that is orthogonal to this line-of-sight. Thus,
  the observer at ${\bf S_1}$ records the stellar velocity to be
  $(V,0)^T$ while observer at ${\bf S_2}$ records the stellar velocity
  to be $(0,V)^T$. The observer at ${\bf S_3}$ however records the
  stellar velocity to have non-zero radial and transverse
  components. Thus, the set of velocities measured by an observer,
  potentially bears information about the observer's location in the
  Galaxy. }
\label{fig:sun}
\end{figure}

\paragraph{}
It may be questioned why the velocity data--observed or
synthetic--alone are invoked to help learn the unknown model parameter
vector $\bX$. Indeed, the data includes information on the spatial
location of the stars as well as the velocity of the stars, but out of
these, only the velocity data can be implemented in the estimation of
$\bX$. This is understood by consulting Figure~\ref{fig:sun}. The
stars that are tracked for their locations and velocities, live in a
circular patch in the neighborhood of the Sun in the 2-dimensional
Milky Way disk; thus, the center of this circular patch is at the
location of the Sun and the radius of this patch is small ($\epsilon$)
compared to $\Vert\bX\Vert$, where $\Vert\cdot\Vert$
is the Euclidean norm of a vector. It is noted that the spatial
location vector $\bp_k$ of the $k$-th star and its velocity vector
$\by_k$, are as recorded by the observer seated at the Sun;
$k=1,2,\ldots,N$ where $N$ stars constitute a data set. Here
$\bp_k=(s_k\cos\alpha_k,s_k\sin\alpha_k)^T$ where $s_k$ is the radial
location of the $k$-th sampled star, as recorded by the heliocentric
observer and $\alpha_k$ is the angular displacement from a chosen
line. The sampling of the spatial locations of the stars is such that
$s_k$ is uniform in the interval $[0,\epsilon]$ and $\alpha_k$ is
uniform in $[0,2\pi]$. Then the mean of $s_k\cos\alpha_k$ over all $k$
is zero as is the mean of $s_k\sin\alpha_k$, i.e. the sample mean of
the measured $\bp_k$ is zero.

\paragraph{}
The left panel of Figure~\ref{fig:sun} shows the location vectors of 3
example stars at points ${\bf P_1}, {\bf P_2}$ and ${\bf P_3}$ inside
this circular patch (marked in grey) where the location of the
observer (Sun) is at point ${\bf S}$ and that of the center of the
Galaxy is at point ${\bf O}$ in the Milky Way disk. Then, in reference
to this figure, the heliocentric location to the $j$-th of these
example stars is the vector ${\bf{SP_j}}={\bf{OP_j}}-{\bf{OS}}$,
$j=1,2,3$ in the figure. But the galactocentric location ${\bf{OP_j}}$
to the star is unknown, implying that the measured heliocentric
location ${\bf{SP_j}}$ cannot be used in this equation to constrain
the location of the observer with respect to the center of the Galaxy,
i.e. the unknown model parameter vector $\bX$ that we are after (or
${\bf{OS}}$ in reference to this figure). Again, the uniform
distribution of $s$ and $\alpha$ suggests that the mean of the
recorded stellar location vectors is zero so that the unknown $\bX$ is
the mean of the galactocentric locations of the sampled stars, which
however is unknown. In other words, no matter what the galactocentric
location of the Sun is, the average of the measured heliocentric
locations of the sampled stars is identically zero; these heliocentric
location measurements do not offer any information about $\bX$.

\paragraph{}
On the other hand, as is depicted in the right panel of
Figure~\ref{fig:sun}, the velocity (vector) of the sampled star at
point ${\bf P}$ as measured by an observer at point ${\bf S_1}$ is
distinct from that measured by the observer at points ${\bf S_2}$ and
${\bf S_3}$ on the Milky Way disk. Here, the velocity of the star
measured by the observer at point ${\bf S_j}$ is considered to have a
radial component along the line ${\bf S_j P}$ that
joins the observer to the star, and the transverse component is
orthogonal to this line; $j=1,2,3$ in this figure. Thus, we see in
this panel that the velocity vector ${\bV}$--that is along ${\textrm{{\bf
      O}{\bf P}}}$--of an example sampled star at ${\bf P}$, will
appear to be entirely along the line ${\bf S_1 P}$
and entirely orthogonal to line ${\bf S_2 P}$, so that its
velocity as measured by the observer at point ${\bf S_1}$ will be
recorded as $(\Vert\bV\Vert, 0)^T$ while the observer at point
${\bf S_2}$ will record its velocity as $(0,
\Vert\bV\Vert)^T$. The observer at point ${\bf S_3}$ will
record the velocity of the star to have non-zero radial and transverse
components. Then, a data set that comprises stellar velocities as
recorded by an observer in the Milky Way disk, bears information about
the location of this observer, i.e. about $\bX$. Thus, such a velocity
data set can be inverted to help estimate the location of the
observer, i.e. the Sun.

\paragraph{}
The radial units used by \citet{chak1} are motivated by the
physics of interaction of the stars in the model Milky Way disk and one of the most conspicuous features in the Galaxy, namely, the elongated stellar bar that rotates with its own rotational frequency $\Omega_b$, pivoted at the center of the Galaxy. The radius at which the (radius-dependent) rotational frequency of the stars in the Milky Way disk equals $\Omega_b$, is called the co-rotation radius or
$R_{CR}$, of that model of the Galaxy. The radial unit used in our work is equivalent to 1$R_{CR}$ for the choices of the Milky Way astrophysical model and $\Omega_b$ used by \citet{chak1}.

\paragraph{}
\citet{chak1} motivates the observer radial location variable to lie in the interval $[1.7,2.3]$ radial units and the observer angular location variable $\theta$ to lie in $[0^\circ,90^\circ]$, on the basis of the relevant physics. These intervals are discretized with $N_R$=24 different values of the radial location and $N_\theta$=9 values of the angular location. The left edge of the radial bin is $r_0$=1.7 radial units, of the angular bin is $\theta_0=0^\circ$, the radial bin width is $\delta_r$=0.025 radial units, the angular bin width is $\delta_\theta$=10$^\circ$. Thus, the $k$-th radial bin is referred to be centered around $1.7 + (k-1)\delta_r + \delta_r/2$, $k=1,2,\ldots,N_R$.
Similarly, the $j$-th angular bin is centered around
$(j-1)\delta_\theta + \delta_\theta/2$, $j=1,2,\ldots,N_\theta$. As
described above, all radial distances expressed here from are in units of
$R_{CR}$ and all angles in units of degrees.

\paragraph{}
The simulations carried out by \citet{chak1} correspond to variation
over the values of the location vector ${\bX}$, the components of
which are the two components of the spatial location vector of the
observer on the Milky Way disk. In these simulations, $\bX$ is chosen
to take values $\bx_1,\bx_2,\ldots, \bx_d$, such that at $\bX=\bx_i$,
the simulated synthetic velocity matrix is $\bY^{(sim)}_i$; here
$i=1,2,\ldots,d$. In these simulations, $d$ was chosen to be 216. Now,
${\bx}_i=(r_i\cos\theta_i,r_i\sin\theta_i)^T$, where the $i$-th value
of the observer radial location is $r_i$ and of the observer angular
location is $\theta_i$.

\paragraph{}
We consider the stellar location and velocity coordinates simulated
from a model of the Milky Way and for each $i\in\{1,2,\ldots,d\}$,
identify the $N_i$ stars that have location vectors such that these
stars lie within a ``neighborhood'' of the $i$-th proposal for the
solar location, i.e. in a ``neighborhood'' of $bx_i$. Here, the size
of the ``neighborhood'' is chosen to mimic the extent of the circular
patch of radius $\epsilon$ centered at the Sun, from within which the
real stars are sampled, to generate the data set $\bY^{(obsvd)}$. The
$i$-th such neighborhood is then defines the intersection of the
$k$-th radial bin and the $j$-th angular bin; $i= N_\theta (k-1) + j$
and in the simulations performed by \citet{chak1}, $\epsilon$
motivated $\delta_\theta$ and $\delta_r$ via the suggestion that
$\pi\epsilon^2$ is roughly approximated by the area of intersection of
a radial and angular bin. The velocity vectors of these $N_i$ stars
are then implemented to estimate the density function from which the
discrete $\bY_i^{(sim)}$ are sampled. This is repeated for each
$i\in\{1,2,\ldots,d\}$. A density function is also estimated from the
real velocity data $\bY^{(obsvd)}$. Pairwise comparison of this
density is undertaken with the density estimated using
$\bY^{(sim)}$. The comparison is parametrized by an affinity parameter
(see Section~\ref{SEC:proposed_method}).
 
\paragraph{}
It merits mention that a set of the synthetic velocity data matrices ${\bY}_i^{(sim)}$, $i=1,\ldots,d$, are obtained with
non-linear dynamical simulations of one, out of four different base
astrophysical models of the Milky Way. However, we do not include any
reference in the notation to the base astrophysical model that the
corresponding synthetic data set is generated from, as we perform
the analysis with each such set of synthetic data, one at a time.
Along with the estimation of the observer, i.e. the solar location in the Milky Way, our investigation aims to determine which of the four astrophysical models best explain the observed data.

\paragraph{}
So to summarize, if $\bX=\bx_{\star}$ represents the location where
the estimated density of the simulated synthetic data has the maximum
affinity with the estimated density of the observed data, our
inference chooses the estimate of the unknown model parameter vector
to be $\bx_{\star}$. We now begin discussion of the details of this
inference that is based on distances between the estimated density of
the observed velocity data at the unknown location and the estimated
density of the synthetic velocity data generated at a chosen value of
$\bX$. Along the way, we will also develop - first the motivation, and
then the methodology used, to implement validation.

\section{Literature Review}\label{SEC:literature}

\paragraph{}
The squared Hellinger distance is one of the most popular measures
used in robust minimum distance inference, and has a one to one
relationship with the Bhattacharyya distance \citep{bhat};
the Hellinger affinity is also referred to as the Bhattacharyya
coefficient. The technical definition of the distance measures,
affinities and coefficients are given in the subsequent
sections. Although it does not satisfy the triangle inequality, the
Bhattacharyya distance is non-negative, and equals zero if and only if
the component densities are identically equal. See \citet{kail, djou, aherne} among others, for
some useful applications of the Bhattacharyya distance in real life
problems. The Hellinger distance is also referred to as the Matusita
distance \citep{matu, kirm} or the Jeffreys-Matusita
distance in the literature. Both the Bhattacharyya distance and the
Matusita (Hellinger) distance (or the corresponding affinities) are
extensively used as measures of separation between probability
densities in many practical problems such as remote sensing
\citep{land, canty}.

\paragraph{}
A method of estimating the solar location in the Milky Way disk, as
proposed by \citet{chak1}, involved performing a $d$ number of tests to test for the null that the
observed data is sampled from the density estimated using the $i$-th
synthetic data set $\bY_i^{(sim)}$ which is generated at the $i$-th
value of the chosen location, i.e. at $\bx_i$ where
$i=1,2,\ldots,d$. The chosen location at which the $p$-value of the test statistic (employed by \citet{chak1}) is maximized, is considered an estimate of the unknown location at which the observed data is realized, i.e. the solar location. In this work, however, our approach is different as we attempt
a direct comparison of the density estimated using the synthetic data $\bY_i^{(sim)}$ and that estimated using the observed data $\bY^{(obsvd)}$, $i=1,2,\ldots,d$. Then $\bx_i$ is our estimated solar location where the closeness of the comparison is quantified by the Hellinger distance. Thus, our work represents an application of the Hellinger distance measure.

\section{Proposed Method}\label{SEC:proposed_method}

\subsection{Motivation}\label{SEC:motivation}

\paragraph{}
Since simulated velocity data at each of the $d$ different chosen locations
are available, the velocity densities at each such point can be
estimated. This we have achieved by fitting a standard bivariate Gaussian kernel. 
Subsequently, we have calculated affinity measures of
each of these densities with the estimated density of the observed
velocity data. The affinity measures so obtained have then been
maximized over the $(r_i, \theta_i)$ grid points to derive the estimate
of the true location from which the observed data may have been
generated.

\paragraph{}
Many choices of density based distances (more generally divergences)
are available in the literature. Depending on their choice, the
distances can exhibit very different characteristics. See \citet{basu} 
for a comprehensive description of the topic of density
based distances and their use in statistical theory. In this
particular work we have chosen to use the affinity measure based on
the Hellinger distance. This affinity measure, also linked to the
Bhattacharya distance, takes the value 1 when the densities coincide,
and takes the value zero when they are singular (i.e. their
supports are non-overlapping). We will give a very brief introduction
to density based distances in the Section~\ref{SEC:distance_measure}.

\subsection{Novelty of the density based method}\label{SEC:novelty}

\paragraph{}
The approach that we adopt in this paper has, in our opinion, the
following advantages to distinguish itself. First of all, we feel
that this is a more natural approach to identify the unknown location
compared to the $p$-value approach \citep{chak1}. The $p$-value
approach for finding the location depends on repeated generation of
data from the physical system or the estimated velocity distribution
for the particular location to create estimates of the
Kullback-Leibler divergence necessary for the generation of the
estimated quantiles for the construction of the $p$-value. We take the
view that since the estimated density for the location is already
available, this is unnecessary. This also greatly reduces the
computational burden of the procedure.

\paragraph{}
Another issue, which has not been sufficiently addressed in the
previous approaches dealing with this problem is the issue of
re-validation of the procedure. Would this method of finding the
maximum of the affinity over the different grid points be
``consistent'' in the sense that should the optimal data generating
location be removed from the data, would the maximum of the affinity
be obtained at one of its immediate neighbors? Essentially we are
demanding a continuity property for the affinity-surface over the
grids in question. We will observe later in the article that in most
situations this is indeed the case for the affinity measure based on
the Hellinger distance, giving us the confidence that the
determination of the location based on the affinity measure is doing
the appropriate thing.

\paragraph{}
There is another point in favor of the particular approach chosen
here. The different models used for the description of the velocity
data-set are, after all, only abstractions of reality. While we expect
that these models will satisfactorily explain the pattern of the
majority of the data, there is always the chance (in fact it is
practically expected) that there could be small subsets of the data
which would not follow the pattern dictated by the bigger majority. In
such situations, the Hellinger distance is a more dependable measure
for identifying the model which fits the large majority of the data,
sacrificing a small group of outlying observations (see, e.g., \citet{basu}). 
The same is not true for the version of the
Kullback-Leibler divergence which generates the $p$-values for the
likelihood based method.

\paragraph{}
Another approach that has been advanced to learn $\bx_{\star}$ is
independent of density estimation \citep{chak3}. In this method, the
velocity data is expressed as a function of the solar location vector
$\bX$ and this unknown function is modeled with a Gaussian
Process. The posterior probability density of $\bx_{\star}$ given the
simulated and observed data is computed in this Bayesian approach. We
compare our results to those obtained by \citet{chak3}.

\subsection{Distance Methods}\label{SEC:distance_measure}

\subsubsection{Affinity Measure based on the Hellinger Distance}\label{SEC:affinity_measure}
\paragraph{}
Let $f$ and $g$ be two probability density functions with respect to the Lebesgue measure (or any other appropriate measure). Then the squared Hellinger distance ${\rm HD}(g, f)$ between the densities $g$ and $f$ is defined as 
\begin{equation}
{\rm HD}(g,f)= \int\left(g^{\frac{1}{2}}(x)-f^{\frac{1}{2}}(x)\right)^{2}dx. \label{EQ:squared_Hellinger_Distance}
\end{equation}
The Hellinger distance is one of the few genuine metrics in the large class of density based divergences widely used in statistics. The measure HD is bounded from above by 2, a value which is attained when the densities are singular. Similarly, the lower bound of the measure is 0, obtained when the densities are identically equal. Notice that the measure in equation (\ref{EQ:squared_Hellinger_Distance}) may be represented as
\begin{align}
\nonumber {\rm HD}(g,f)  &= \int g(x) dx + \int f(x)dx - 2 \int g^{\frac{1}{2}}(x)f^{\frac{1}{2}}(x)dx \\ 
& = 2\left(1 - \int g^{\frac{1}{2}}(x)f^{\frac{1}{2}}(x)dx \right). \label{EQ:squared_Hellinger_Distance_representation}
\end{align}
Thus the minimization of the Hellinger distance is equivalent to the maximization of the affinity measure 
\begin{equation}\label{EQ:Hellinger_affinity}
\rho(g,f)=\int g^{\frac{1}{2}}(x)f^{\frac{1}{2}}(x)dx
\end{equation} which varies between 0 and 1; the end points are obtained when the the densities are singular  and identical respectively. The quantity in 
(\ref{EQ:Hellinger_affinity}) is linked to the Bhattacharyya distance \citep{bhat}
\begin{equation}
B(g, f) = -\log \left(\int g^{\frac{1}{2}}(x)f^{\frac{1}{2}}(x)dx\right),
\end{equation}
and is widely used as a measure of closeness between two probability densities.

\subsubsection{The Kullback-Leibler (KL) Divergence}\label{SEC:KL_div}

\paragraph{}
The Kullback-Leibler divergence (also known as information divergence, information gain, relative entropy) is a non-symmetric 
measure of the difference between two probability distributions $G$ and $F$ \citep{kull}. The distribution $G$ typically represents the ``true" distribution of the data while the distribution $F$ represents a theory, model, description, or approximation of $G$.

\paragraph{}
Although it is often intuited as a metric or distance, the KL divergence is not a true metric. In particular it is not a symmetric measure; the KL divergence between $G$ and $F$ is  generally not the same as that between $F$ and $G$. The divergence is computed between the corresponding densities $g$ and $f$, and is defined as:
\begin{equation}\label{EQ:Kullback-Liebler}
\delta(g,f)=\int g(x)\log\left(\frac{g(x)}{f(x)} \right)dx.
\end{equation}
This divergence measure is not bounded above; however, a zero value of this measure indicates zero distance between $f$ and $g$, i.e. the densities are identically equal.  
In spirit, the divergence measure can be considered to be similar to the inverse of the affinity measure. Both the KL and HD measures are special cases of the Cressie-Read family of power-divergences \citep{cres}.
\subsubsection{Relative Pearson (rPE) divergence}\label{SEC:rPE_div}
The Pearson (PE) divergence is a squared-loss variant of the Kullback-Leibler divergence. It is basically an extension of the Pearson's $\chi^2$ divergence and is defined as
\begin{equation}\label{EQ:Pearson Divergence}
PE(g,f)= \int g(x)\left(\frac{f(x)}{g(x)}-1 \right)^2dx.
\end{equation}
It also belongs to the family of $f$-divergences and share many theoretical properties of the KL Divergence. This divergence measure is also not bounded above; a zero value of this measure indicates zero distance between $f$ and $g$, i.e. the densities are identically equal.

The relative Pearson divergence (rPED) is a variant of the Pearson divergence (see, e.g., \citet{sugi}). It is defined as
\begin{equation}\label{EQ:Relative Pearson Divergence}
rPE(g,f)= PE(h_\alpha,f)=\int h_\alpha(x)\left(\frac{f(x)}{h_\alpha(x)}-1 \right)^2dx,
\end{equation}
where
\begin{equation}\label{h_alpha}
h_\alpha(x)=\alpha f(x)+(1-\alpha)g(x)\,\,\,\,\,\text{ for }\,\,\,\,0\leqslant \alpha<1.
\end{equation}

For $\alpha=0$, the relative Pearson divergence reduces to the normal PE divergence. However, the relative density ratio in this case, i.e. $f/h_\alpha$, is bounded above by $1/\alpha$ for $\alpha>0$:
\[
\frac{f(x)}{h_\alpha(x)}=\frac{1}{\alpha+(1-\alpha)\frac{g(x)}{f(x)}}<\frac{1}{\alpha}.
\]
Thus it overcomes the problem of the unboundedness of the density ratio $f/g$ in the PE divergence. The tuning parameter $\alpha$ is chosen by cross-validation.
\section{Results}\label{SEC:results}

\paragraph{}
The four astrophysical models will henceforth be referred to as 18sp3bar3, 25sp3bar3, 
sp3bar3 and bar6. This nomenclature involves the values of the bar and the spiral parameters 
which specify the models.

\subsection{Density Estimation}\label{SEC:density_est}

\paragraph{}
We use the bivariate kernel density estimation with the kernel $K(\cdot,\cdot)$ being the 
standard two dimensional Gaussian kernel with covariance matrix $\bI_2$, the two dimensional 
identity matrix, i.e. the kernel function is given by 
\begin{equation}
	K(x,y)= \frac{1}{2\pi}\exp\left(-\frac{x^2+y^2}{2} \right).
\end{equation}
Based on a set of $n$ independent and identically distributed observations $(X_1,Y_1), \hdots,
(X_n,Y_n)$ from the data generating density, our density estimate is given by 
\begin{equation}
	\hat{f}(x,y)=\frac{1}{nh^2}\sum\limits_{i=1}^n K\left(\frac{x-X_i}{h},\frac{y-Y_i}{h}\right),
\end{equation}
where $h$ is the smoothing parameter. We have chosen the smoothing parameter $h$ as 
\begin{equation}
	h =\sigma n^{-\frac{1}{6}},
\end{equation}
where
\[
	\sigma^2=\frac{s_X^2+s_Y^2}{2}.
\]
Here $s_X^2$ and $s_Y^2$ are the sample variances of the $X$  and the $Y$ observations respectively. See, e.g., \citet{silv} for a discussion on the choice of the smoothing parameter. 

\paragraph{}
For a fixed model, let us denote the true density at the location $(r,\theta)$ under this model by $g_{(r,\theta)}(x,y)$ and its kernel density estimate by $\hat{g}_{(r,\theta)}(x,y)$. Also we shall denote the true density of the observed velocity data by $f(x,y)$ and its kernel density estimate by $\hat{f}(x,y)$. As the analysis for the simulated data generated at the $d=216$ chosen locations each for each of the base astrophysical models is done separately, we do not attach another index for this base model to the density $g(\cdot,\cdot)$.

Here the observed velocity vectors are assumed to be independent and identically distributed. Such assumptions are generally reasonable and frequently employed in astronomical studies. See, e.g., \citet{babu,way}.
\subsection{Maximum Affinity Estimation of the Location Parameter}\label{SEC:affinity}

\begin{figure}[htbp]
\includegraphics[scale=0.6]{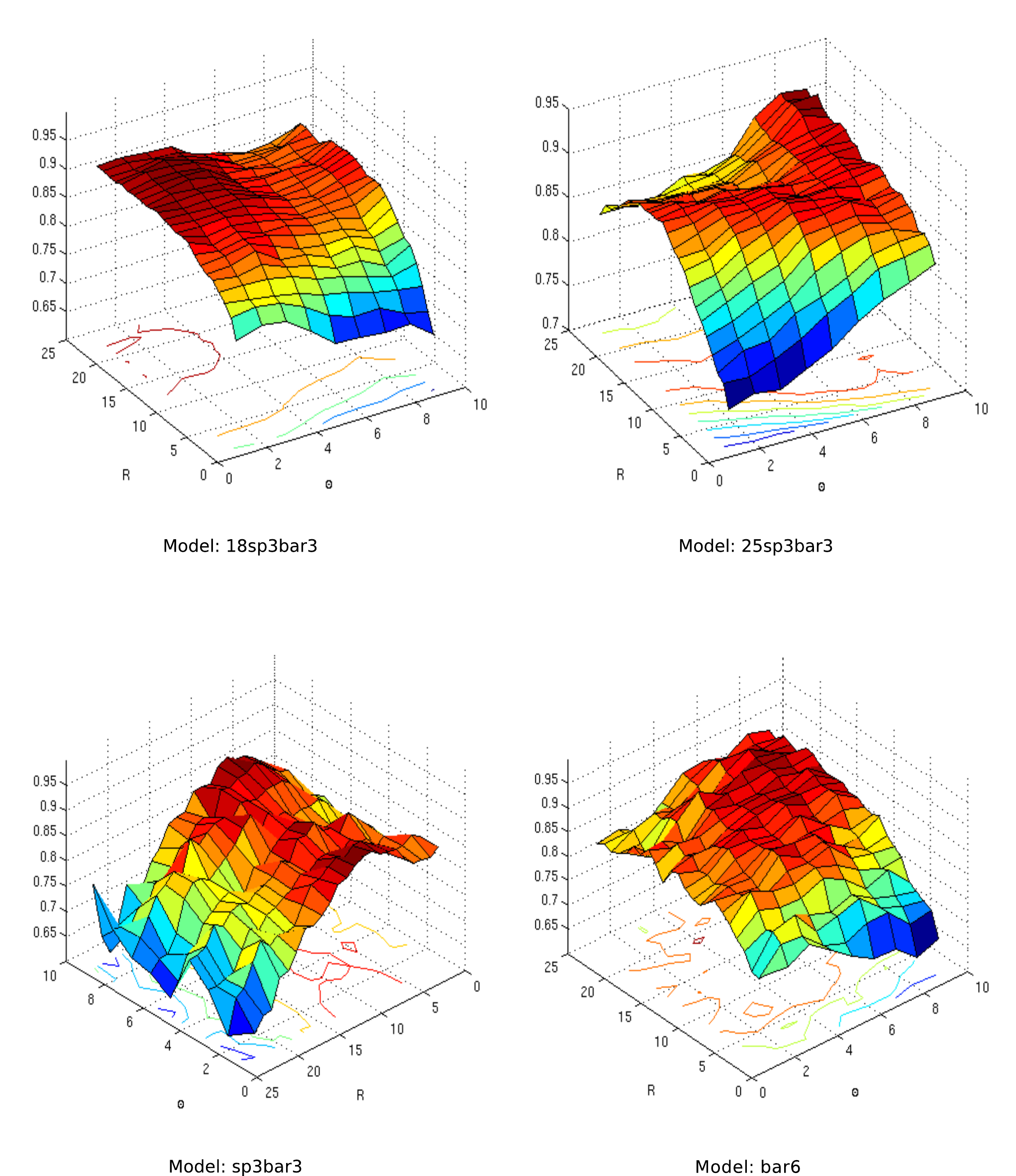}
\caption{Hellinger affinity surfaces under the four different base astrophysical models.}
\label{FIG:affinity_surfaces}
\end{figure}

\paragraph{}
Here we present the results of the proposed method for each of the four simulation models. For a given model, let us define
\begin{equation}
\rho_{(r,\theta)}:=\rho(g_{(r,\theta)},f).
\end{equation}
where $\rho(\cdot,\cdot)$ is the Hellinger affinity defined in Equation (\ref{EQ:Hellinger_affinity}). 
We compute the density estimate $\hat{f}$ for the observed data $\bY^{(obsvd)}$. We also compute the density $\hat{g}_{(r_i,\theta_i)}$ for the synthetic data $\bY_i^{(sim)}$ that is simulated at the $i$-th chosen location $(r_i\cos\theta_i,r_i\sin\theta_i)^T$ out of the $d=216$ such chosen locations. We use the former and latter density estimates as surrogates for $f$  and $g_{(r,\theta)}$ respectively. Thus for each base astrophysical model, we have 216 affinity values corresponding to each $i$; for brevity's sake, we use the notation
\begin{equation}
	\hat{\rho}_{(r_i,\theta_i)}=\rho(\hat{g}_{(r_i,\theta_i)},\hat{f}).
\end{equation}
In Figure \ref{FIG:affinity_surfaces}, we show the affinity surfaces
generated over the chosen locations at which the simulated velocity
data are generated in each of the base astrophysical models, i.e. the
surface plot of $\hat{\rho}_{(r_i,\theta_i)}$ against $(r_i,\theta_i)$
for $i=1,2,\ldots,d$, $d=216$, for each base model. The plot provides
visualization of where the surfaces attain their maxima.

\begin{figure}[htbp]
\includegraphics[scale=0.5]{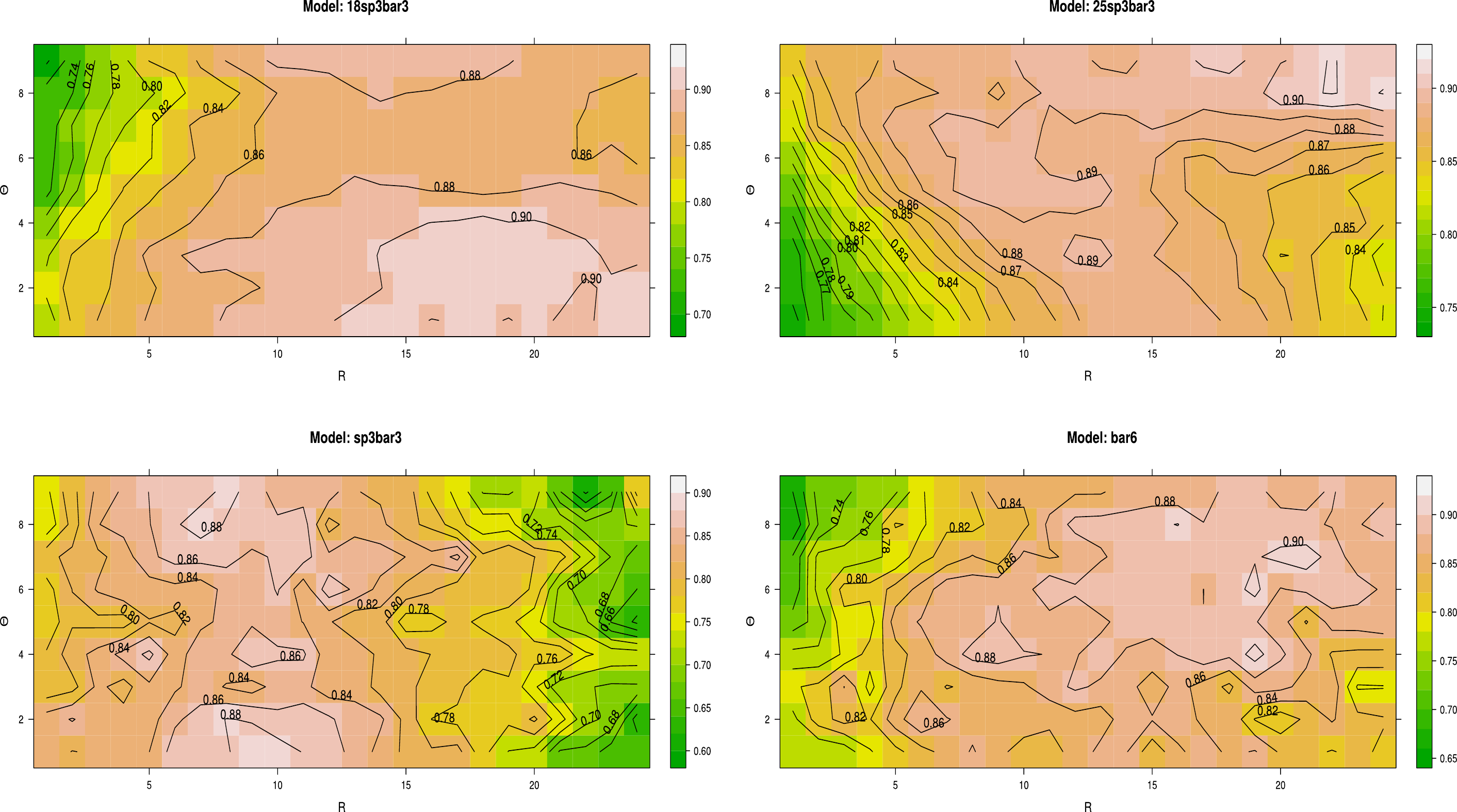}
\caption{Contour plots of the Hellinger affinity surfaces under the four different base astrophysical models.}
\label{FIG:affinity_contours}
\end{figure}

\paragraph{}
To avoid the dependency on perspective while viewing a surface plot,
it sometimes helps to look at a more mundane contour plot. Figure
\ref{FIG:affinity_contours} shows a color mapped contour plot of the
affinity surface. It is clear from these contour plots that in the
bar6 model there is a single mode, while in the sp3bar3 model there
are at least two pronounced modes. The other two models fall somewhere
in between. This is in concert with the results obtained earlier by
\citet{chak2}, and also by \citet{chak3}.

\paragraph{}
The chosen locations at which the synthetic data are simulated from the base astrophysical model in question are in fact arranged over a uniform rectangular grid. Thus, the $i$-th point in this grid would represent the $i$-th such chosen location, $i=1,2,\ldots,d$, $d=216$ as per the non-linear dynamical simulations of the Milky Way disk reported in \citet{chak1}. Taking advantage of the uniform nature of this grid, any grid point could have an alternative, 2-dimensional representation: $(k, j)$, $k = 1, 2,\ldots, 24$, $j = 1, \ldots, 9$, so that there are $24\times 9=216$ such chosen locations. In this treatment, let the $(k,j)$-th grid point be the physical location $(r_k,\theta_j)$.

\paragraph{}
For the base astrophysical model in question, we define $\max(k, j)$ as the indices for 
the particular chosen location where the affinity measure is maximized. Let the corresponding 
radial and angular coordinates of this location be
\begin{equation}
(r_{\max},\theta_{\max}):=\arg\max\limits_{(k,j)}\rho(g_{(r_k,\theta_j)},f).
\end{equation}
In words, $(r_{\max},\theta_{\max})$ is the actual physical location
where the true distribution of the synthetic data is closest to the true
distribution of the observed data in the sense of having highest
affinity, while $\max(k, j)$ represents the indices for this
location. We have estimated $(r_{\max},\theta_{\max})$ by
\begin{equation}
(\hat{r}_{\max},\hat{\theta}_{\max})=\arg\max\limits_{(k,j)}\hat{\rho}_{(r_k,\theta_j)},
\end{equation}
and the corresponding indices provide an estimate of $\max(k,j)$. We refer to this estimate 
of $\max(k,j)$ as $\widehat{\max(k,j)}$. 

\paragraph{}
Out of the chosen 216 grid points, the location
$(\hat{r}_{\max},\hat{\theta}_{\max})$ corresponding to the
$\widehat{\max(k,j)}$-th grid point is the one that that maximizes the
affinity of the density of the observed data to that of the simulated
data. In other words, out of the 216 chosen locations in our work,
this location best represents the value of the unknown model parameter
vector $\bX$ at which the observed data $\bY^{(obsvd)}$ are
realized. Since $\bX$ is the unknown location of the observer who
observes data $\bY^{(obsvd)}$, $\hat{r}_{\max}$ and
$\hat{\theta}_{\max}$ best represent the values of the unknown radial
and angular location of the observer respectively, out of the set of
chosen locations that we use in our work, following the
astrophysically motivated choice of such parameters by \citet{chak1}.

\paragraph{}
In Table \ref{table1} we present the estimated locations (and their indices) where the affinities are maximized for the four base astrophysical models.

\begin{table}[htbp]
\caption{Location of maximum affinities for the four base astrophysical models.}
\label{table1}
\begin{center}
\begin{tabular}{ccc}\hline
 Model & $\widehat{\max(i, j)}$ & $(\hat{r}_{\max},\hat{\theta}_{\max})$\\ \hline\hline
 18sp3bar3 & (20, 2) & (2.1875, 15$^\circ$)\\ \hline
 25sp3bar3 & (22, 9) & (2.2325, 85$^\circ$)\\ \hline
 sp3bar3 & (10, 1) & (1.9375, 5$^\circ$)\\ \hline
 bar6 & (21, 7) & (2.2125, 65$^\circ$)\\ \hline
\end{tabular}
\end{center}
\end{table}

Thus the location of the maximum affinities are quite different for
the four models. Note that these point estimates are not going to be very
precise owing to the multimodal and flat character of the affinity
surfaces. The contour plots in Figure \ref{FIG:affinity_contours}
provide more meaningful information. In Section~\ref{SEC:confidence} we shall provide
confidence sets around these point estimates and in light of those results, will carry out a comparison of our results to those reported by \citet{chak1} and \citet{chak3}.

\subsection{Maximum Entropy Estimation of the Location Parameter}\label{SEC:KLD}

\begin{figure}[htbp]
\includegraphics[scale=0.7]{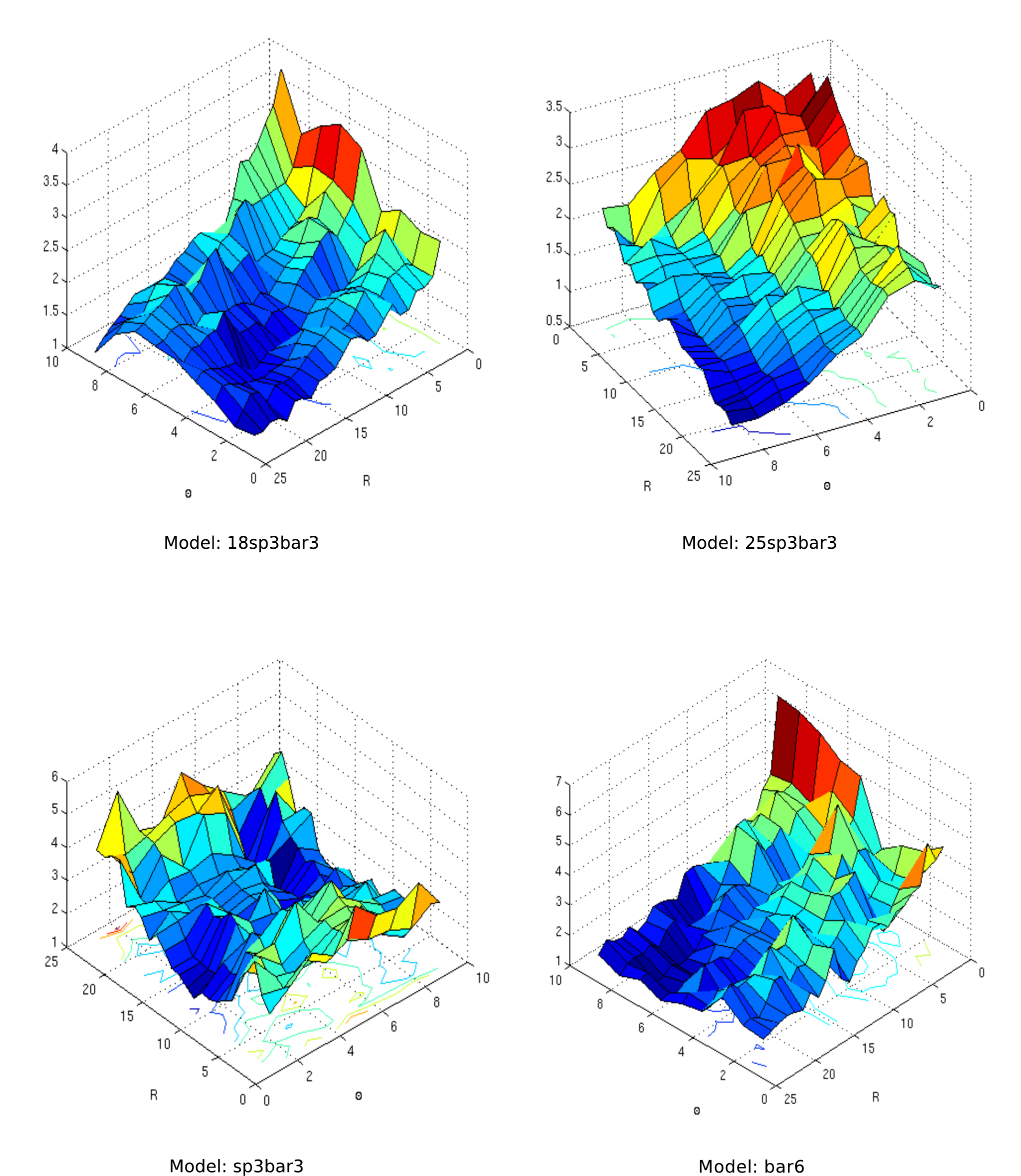}
\caption{KLD surfaces under the four different base astrophysical models.}
\label{FIG:KLD_surfaces}
\end{figure}

\paragraph{}
For a given base model, we define
\begin{equation}
\delta_{(r,\theta)}:=\delta(g_{(r,\theta)},f),
\end{equation}
where $\delta(\cdot,\cdot)$ is the Kullback-Liebler Divergence (KLD) defined in Equation (\ref{EQ:Kullback-Liebler}). 
Let the density estimate for the observed data $\bY^{(obsvd)}$ be abbreviated as $\hat{f}$, and the estimated density of the data $\bY_i^{(sim)}$ simulated at the chosen location $(r_i\cos\theta_i,r_i\sin\theta_i)^T$ be $\hat{g}_{(r_i,\theta_i)}$, $i=1,2,\ldots,d$, $d=216$. Thus for each base astrophysical model we have 216 KLD values at each of the 216 chosen locations at which the synthetic data sets are generated. For simplicity of notation we again denote
\begin{equation}
	\hat{\delta}_{(r_i,\theta_i)}=\delta(\hat{g}_{(r_i,\theta_i)},\hat{f}).
\end{equation}
In Figure \ref{FIG:KLD_surfaces}, we show the KLD surface generated at the 216 chosen locations for each of the base astrophysical models, i.e. the surface plot of $\hat{\delta}_{(r_i,\theta_i)}$ against $(r_i,\theta_i)$. The plot helps to visually detect where the surfaces attain their minima. 

\begin{figure}[htbp]
\includegraphics[scale=0.5]{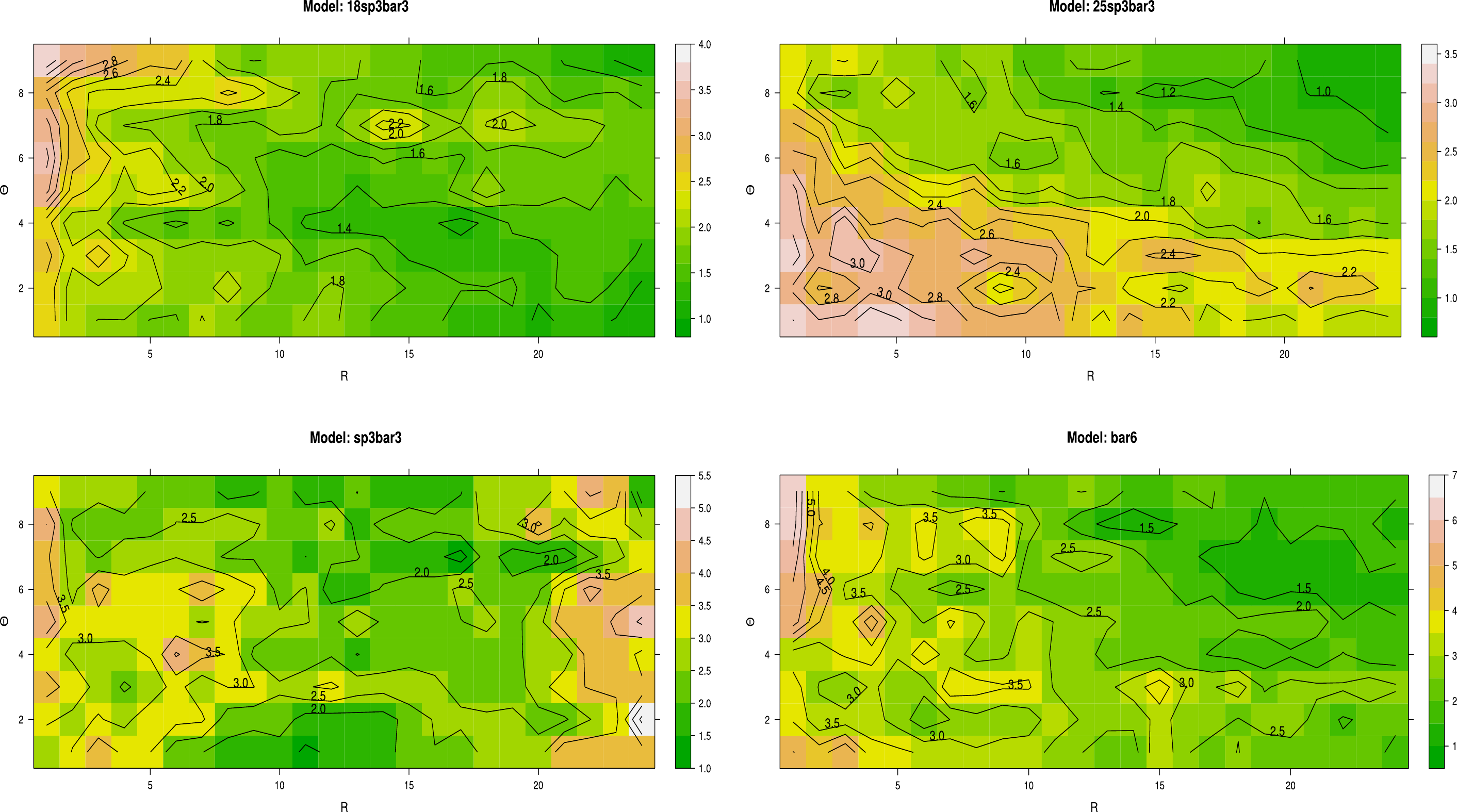}
\caption{Contour plots of KLD surfaces under the four different astrophysical models.}
\label{FIG:KLD_contours}
\end{figure}

\paragraph{}
As with the affinity plots, here too we display the color mapped KLD contours; see Figure \ref{FIG:KLD_contours}. Note that the overall appearance of these contour plots is in agreement with Figure \ref{FIG:affinity_contours}.

\paragraph{}
Again, as in the discussion of Section~\ref{SEC:affinity}, here too we
invoke the construct that the $d$ (=216) chosen locations are placed
on a uniform 2-dimensional rectangular grid. Then each grid point can
be represented by a pair of indices such as $(k,j)$, $k=1,2,\ldots,24$,
$j=1,2,\ldots,9$. The location of the $(k,j)$-th grid point is
$(r_k,\theta_j)$. Let $\min(k, j)$ represent the indices for the
particular grid point where the KLD values are minimized with the physical
location of this grid point represented by 
\begin{equation}
(r_{\min},\theta_{\min})=\arg\min\limits_{(k,j)}\delta(g_{(r_k,\theta_j)},f).
\end{equation}
Thus, $(r_{\min},\theta_{\min})$ is the actual physical location where
the true distribution of the simulated data is closest to the true
distribution of the observed data in the sense of having lowest KLD,
while $\min(k, j)$ represents the indices for this location. We
estimate this location $(r_{\textrm min},\theta_{\textrm min})$ by
\begin{equation}
(\hat{r}_{\min},\hat{\theta}_{\min})=\arg\min\limits_{(k,j)}\hat{\delta}_{(r_k,\theta_j)},
\end{equation}
and the corresponding indices provide an estimate of $\min(k,j)$.

\paragraph{}
In Table \ref{table2} we present the coordinates of the location where the KLD values are minimized for the four base astrophysical models.
\begin{table}[htbp]
\caption{Location of minimum KLD for the four models}
\label{table2}
\begin{center}
\begin{tabular}{ccc}\hline
 Model & $\widehat{\min(k, j)}$ & $(\hat{r}_{\min},\hat{\theta}_{\min})$\\ \hline\hline
 18sp3bar3 & (24, 9) & (2.2875, 85$^\circ$)\\ \hline
 25sp3bar3 & (24, 8) & (2.2875, 75$^\circ$)\\ \hline
 sp3bar3 & (17, 7) & (2.1125, 65$^\circ$)\\ \hline
 bar6 & (22, 7) & (2.2375, 65$^\circ$)\\ \hline
\end{tabular}
\end{center}
\end{table}

\begin{figure}[htbp]
\includegraphics[height=18cm,width=\textwidth]{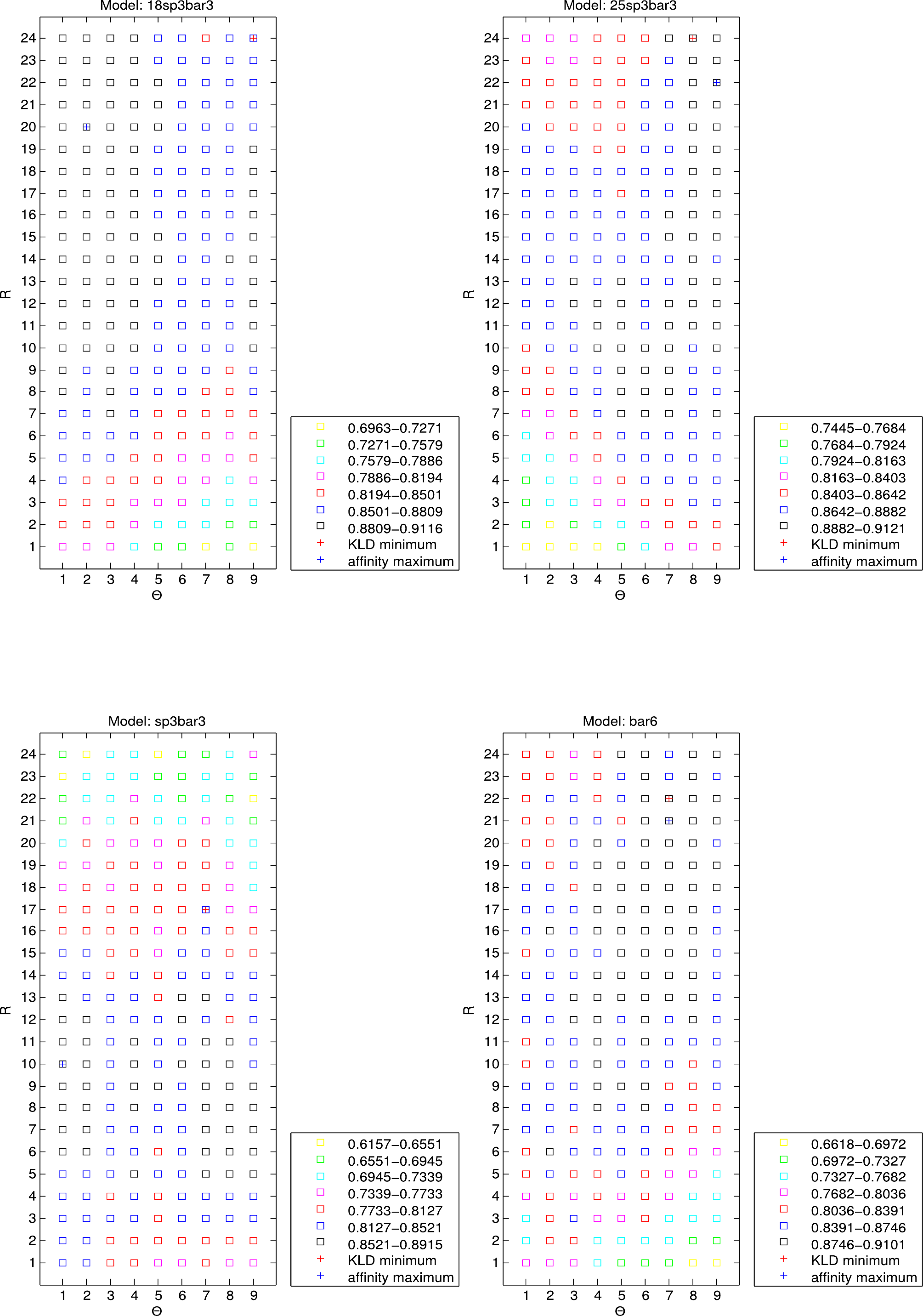}
\caption{A discrete representation of the level-plots of the affinity measure recovered in the
  2-dimensional grid of our chosen locations. Locations at which
  values of the recovered affinity measure lie in the same band, are
  marked in the same color. The color coding of the affinity measure
  values is presented in the key adjoining each panel. }
\label{FIG:aff_comp_aff_KLD}
\end{figure}

\begin{figure}[htbp]
\includegraphics[height=18cm,width=\textwidth]{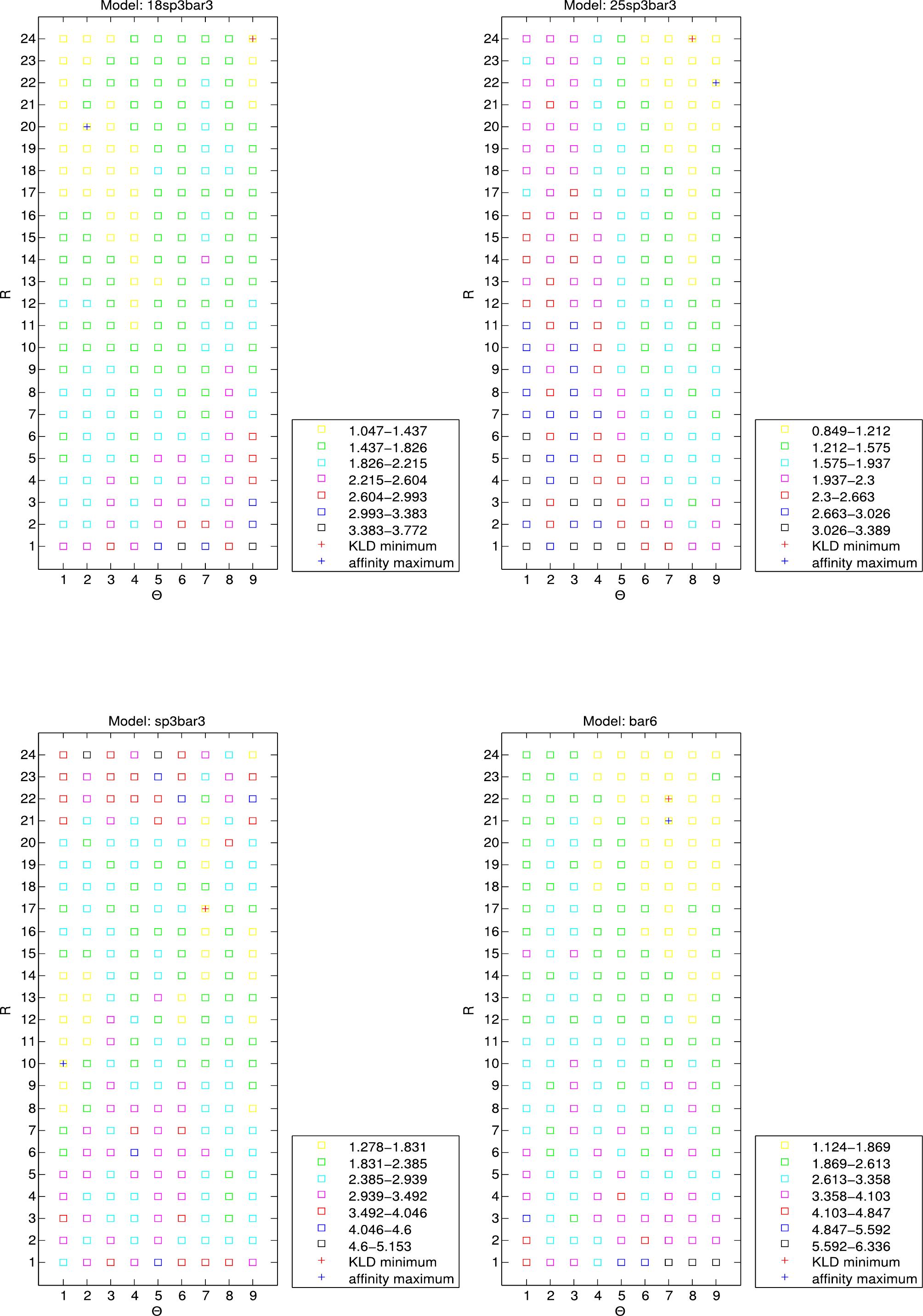}
\caption{Level-plots of the KLD surface, the analogous plot to Figure~\ref{FIG:aff_comp_aff_KLD}.}
\label{FIG:KLD_comp_aff_KLD}
\end{figure}

\paragraph{}
Figure \ref{FIG:aff_comp_aff_KLD} shows level-plots of the affinity
surface along with the KLD estimates. A corresponding KLD version is
shown in Figure \ref{FIG:KLD_comp_aff_KLD}. Note that the estimates
provided by these two approaches are quite close.


\paragraph{}
It is interesting to note that the surfaces are quite flat
(particularly in the case of the base models ``18sp3bar3'' and
``bar6'') near the peaks. Therefore, estimation of the location for
which the affinity attains the maximum becomes difficult. One needs to
investigate further to see whether this method will produce the right
location ``consistently''. 

\paragraph{}
In particular, we are concerned that method of
estimation used in or work abides by the undertaken assumptions.  To
this effect, we seek validation of our results.

\paragraph{}
At the same time, we are interested in quantifying uncertainties in
the estimated locations out of the chosen $d$ locations at which
the density of the synthetic data approaches the density of the
observed data closest, in the sense that the affinity measure between
this pair of densities is the highest. In order to perform parameter
uncertainty estimation, we undertake the construction of confidence
sets using a bootstrap based method.
\subsection{Confidence Sets}\label{SEC:confidence}

\paragraph{}
We are interested in quantifying uncertainties in the estimation of
the locations (inside the grid of our choice) at which the affinity
measure in maximized. We recall that $(r_{\max},\theta_{\max})$ is the
location at which the true distribution under the model is closest to
the true distribution of the observed data in the sense of having
highest affinity among densities. We generated 300 bootstrap samples
from the density of the synthetic data generated at
$(r_{\max},\theta_{\max})$.  We then computed the affinity measures
between the true density of the observed data and the bootstrap
samples. This gave rise to a sampling distribution of the affinity
measures between the density of the observed data and the bootstrap
samples from the $(r_{\max},\theta_{\max})$ location.  Locations at
which the values of the affinity measures
(i.e. $\hat{\rho}(r_i,\theta_{i})$) are above the cut-off point were
included in the confidence set. For a $95\%$ confidence set, we chose
the lower fifth percentile of the empirical affinity distribution
obtained through the above described bootstrap exercise as the cut-off
point. However, we acknowledge that the suggested confidence sets will
be valid under the assumption that the contours of constant affinity
are shift invariant as $(r_{\max},\theta_{\max})$ is varied.

\begin{figure}[htbp]
\centering
\includegraphics[height=18cm,width=0.95\textwidth]{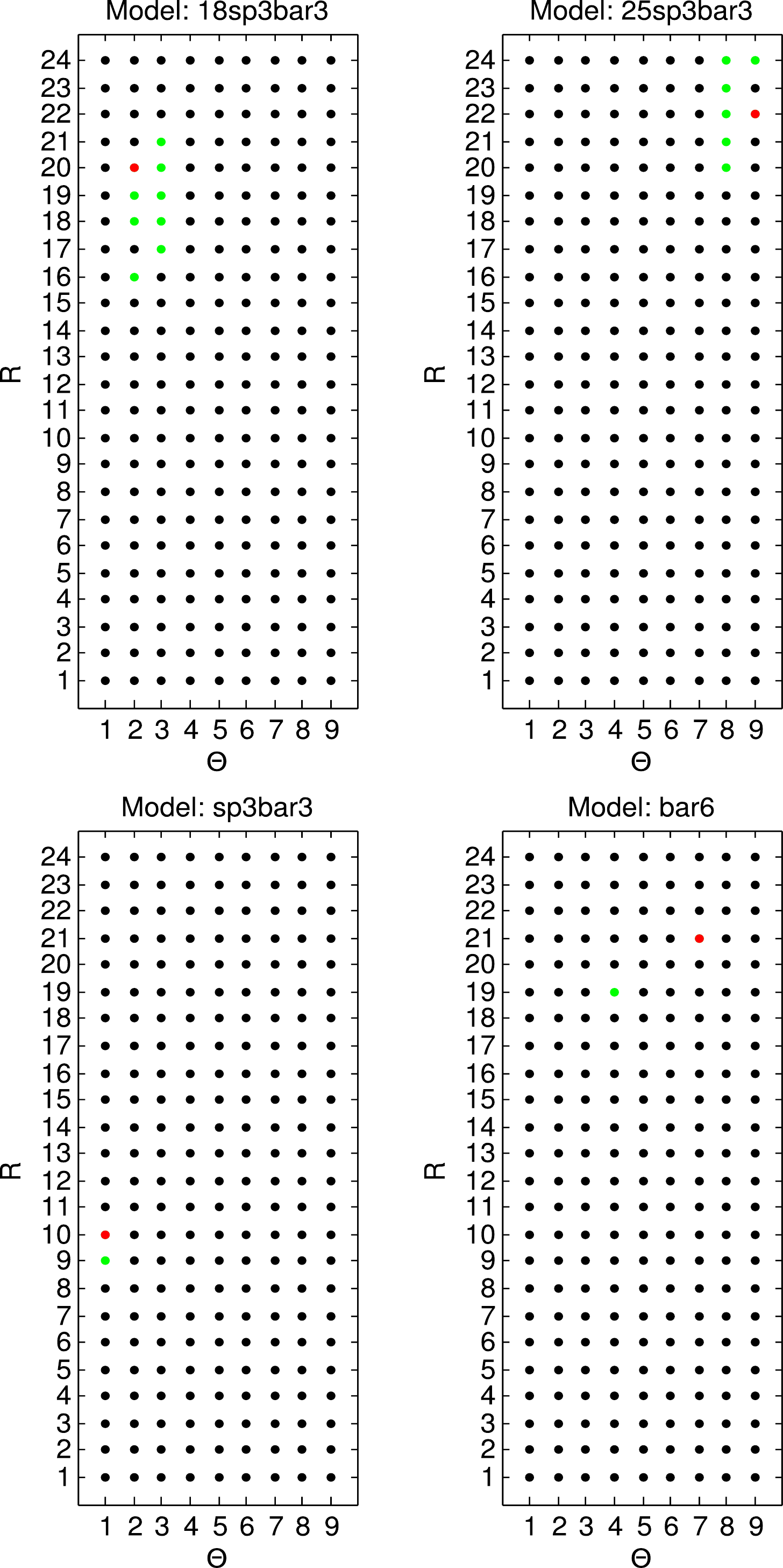}
\caption{$95\%$ confidence set for $(r_{\max},\theta_{\max})$ under each model. 
The elements of the confidence sets are depicted as green dots, the red dots 
representing the point-estimates obtained earlier.}
\label{FIG:conf_fig}
\end{figure}

\paragraph{}
In Figure \ref{FIG:conf_fig} we show the confidence sets. The actual point where 
the affinity is maximized is indicated in red, while the other points in the 
confidence set are indicated in green. It is interesting that 
the confidence sets for all the models are fairly small, and for the last two 
models the sets have just two members each. This shows that the estimation procedure 
is quite precise. 

\paragraph{}
These estimates overlap moderately well with those reported by
\citet{chak1} as well as those by \citet{chak3}.  For the base
astrophysical model bar6, \citet{chak1} reports that the angular
location of the Sun lies between 0$^\circ$ and 49$^\circ$ with a
median at 22$^\circ$ while the radial location$\in$[1.9625,
  2.1975]. For this model, \citet{chak3} suggests that the mode of the
marginal posterior probability density of $r_\star$ occurs at 2.2 and
of $\theta_\star$ at 23.5$^\circ$. In this Bayesian estimate of
\citet{chak3}, the estimates lie in 95$\%$ highest probability density
(HPD) credible regions that are respectively $[2.04,2.3]$ and about
$[21^\circ,26^\circ]$. As evident in Table \ref{table1}, our point
estimate for this base model is too high to fit into this
interval. However, the confidence set estimated for this base model
includes locations at lower values of the radial location as well as
lower angular location values (shown in green in Figure
\ref{FIG:conf_fig}), such that these values are in conformity with the
findings of \citet{chak1} and \citet{chak3}.

\paragraph{}
For the base astrophysical model 18sp3bar3, the radial and angular
location estimates of \citet{chak1} are [1.95, 2.21] and [0$^\circ$,
  30$^\circ$] respectively. The estimates of \citet{chak3} are
similar, with the 95$\%$ HPD credible region given by [1.7, 2.29] and
about [10$^\circ$, 62$^\circ$] for the solar radial and angular
coordinates respectively. Our point estimate of (2.1875, 15$^\circ$)
for this base model then lies comfortably within these intervals; the
confidence set recovered for this base model suggests that the
observed data are consistent with radial location values lower than
2.1875 at the angular location value of 15$^\circ$ as well as at a
higher angular value of 25$^\circ$. In fact, a slightly higher radial
location value of 2.2125, at an angular value of 25$^\circ$ is also
included in our constructed confidence set for this base model.

\paragraph{}
This example helps to bring to the fore a salient advantage of the
uncertainty estimation in our work, compared to that in \citet{chak1}. 
Given that we are performing a joint (radial and angular)
parameter uncertainty estimation, we present our results as confidence
sets on the 2-dimensional grid of our chosen locations. This allows
for identification of the interval estimate of the solar location more
clearly than in \citet{chak1} in which the intervals represent
uncertainties on the radial or angular values obtained using the
marginal distribution of the radial or angular location values. Thus,
it needs to be emphasized that the interval estimation of \citet{chak1} 
are not to be directly compared to our estimated uncertainties.
Additionally, the 95$\%$ HPD credible regions that \citet{chak3} report
are fundamentally different from our uncertainty estimates.
We merely explore the possibility of an overall overlap between the
results obtained using our methodology here with what exists in the
literature.

\paragraph{}
In the context of our uncertainty estimation, we would also like to
emphasis that it has been discussed in the literature that the
underlying chaos in the base astrophysical model drives the estimated
locations to be scattered over the constructed grid of the chosen
locations \citep{chak1, chak2}. In fact, a necessary condition for chaos to occur is the increasing non-injectivity of stellar velocity as a function of the unknown solar location $\bX$ \citep{sengupta}. In the results presented by
\citet{chak3}, the models that manifest such chaos are those for which
the posterior probability density of the location parameters are
rendered multimodal. In other words, the distribution of the locations
that are compatible with the observed data (i.e. the locations
at which the affinity measure is high in our work), may be
multimodal. This further suggests that a visual representation of the
confidence sets (as in Figure \ref{FIG:conf_fig}) allows for easy
reading of the interval estimation of the unknown solar location.

\paragraph{}
Of all the base models, \citet{chak1} had found the distribution of
locations compatible with the observed data to be most scattered over
the grid of chosen locations for sp3bar3. This scatter disallowed the
interval estimation of the unknown solar location in this earlier
work. \citet{chak3} agrees with this trend in that the posterior
densities of the location parameters are most multimodal for this
model. We confirm a similar trend in our recovery of the affinity
surfaces (Figure \ref{FIG:aff_comp_aff_KLD}). However, our method of
estimating uncertainties works for this base model and we recover a
very small confidence set adjoining the point estimate at (1.9375,
5$^\circ$); see Figure \ref{FIG:conf_fig}.

\paragraph{}
For the base model 25sp3bar3, our point estimate equals (2.2325,
85$^\circ$) (see Table \ref{table1}) while the recovered confidence
set suggests that at a slightly lower angular location value of
75$^\circ$, radial location values in [2.1875, 2.2875] are also
compatible with the observed data as they are within the 95$\%$
confidence interval; the situation is the same for the location 2.2875
at the higher angle of 85$^\circ$.  While these radial location values
overlap with the estimate from \citet{chak1}, our estimate of the
angular locations are slightly in excess of the earlier estimate of
angular location value.

\subsection{Cross-validation}\label{SEC:validation}

\paragraph{}
In this context it is recalled that the estimation procedure is based
on the assumption that the velocities observed from nearby locations 
and hence their corresponding densities will be more similar to each
other than those observed from distant locations. It is important to
verify that the affinity values obtained by this method show such
desirable property. For this purpose, we used a cross-validation
approach where one of the grid points was chosen as the ``true''
location, and the corresponding kernel density estimate was chosen as
the ``true'' density. The affinity values between this density and the
density estimates at all the other grid points are obtained. Under the
aforementioned assumption it is expected that the maximum of these
affinity values should occur at one of the nearest neighbors of the
``true'' locations. For this analysis, the $24 \times 9$-sized grid 
$(r_k, \theta_j), k=1,\ldots, N_R, j=1,\ldots, N_\theta$ was broken into 24
blocks of size $3 \times 3$ each. The mid-point of each block was
chosen as the representative for that block for the purpose of cross-validation; 
thus for each base
astrophysical model, we had 24 points implemented in cross-validation.

\begin{figure}[htbp]
\includegraphics[scale=0.5]{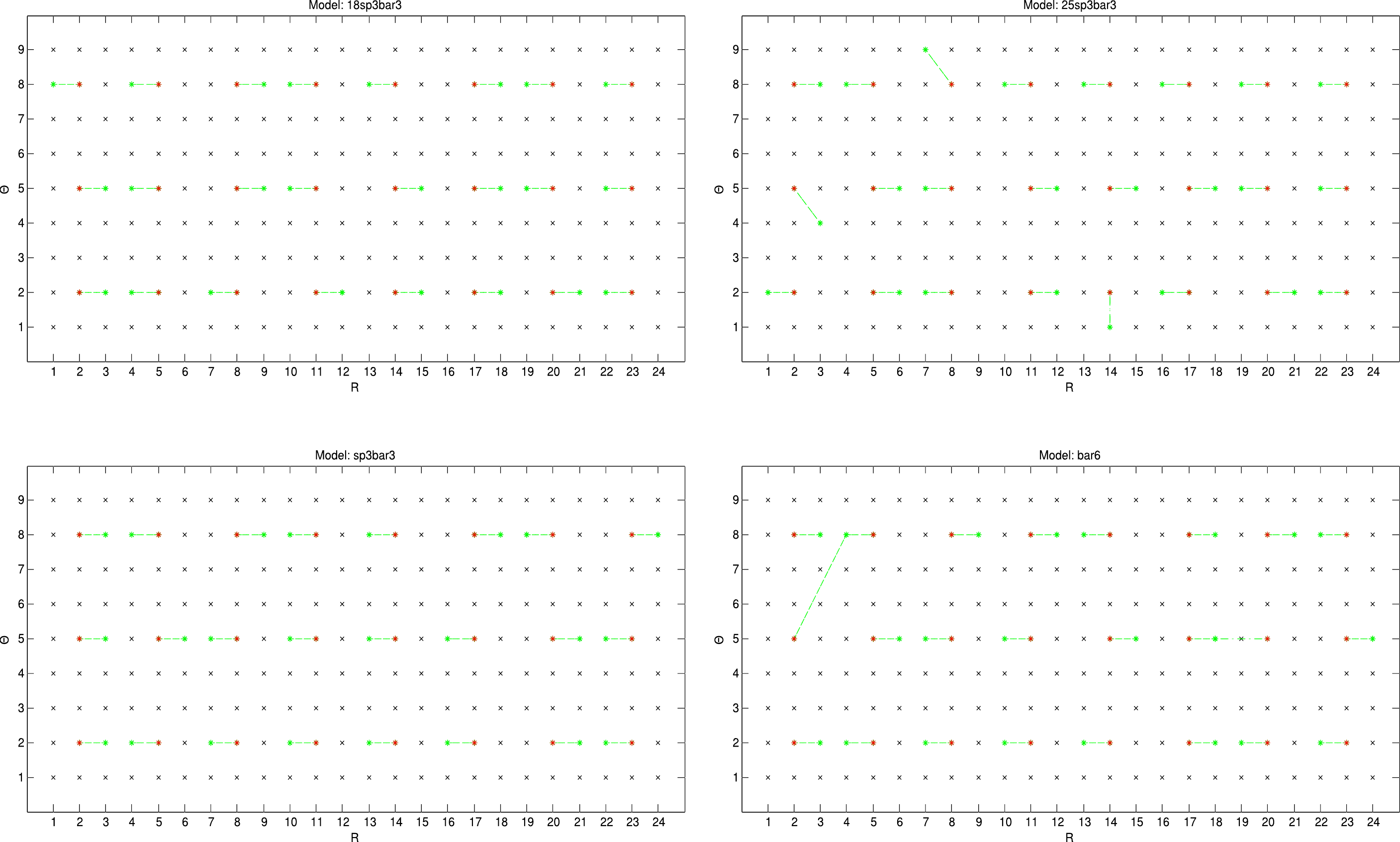}
\caption{Graph depicting the locations where the maxima occurred during cross-validation. The 
red stars represent the co-ordinates chosen for cross-validation and green lines connect these 
co-ordinates to their corresponding maxima (represented by green stars).}
\label{FIG:cv_graph}
\end{figure}

\paragraph{}
When the midpoint $(r_{k_m}, \theta_{j_n})$ is chosen as the true
location, we define its first neighborhood points as the set of points
$(r_k, \theta_j)$ such that $\max(|k-k_{m}|, |j-j_{n}|) =1$,
its second neighborhood points as the set of points $(r_k, \theta_j)$
such that $\max(|k-k_{m}|, |j-j_{n}|) =2$ and so on.
In Table \ref{table3}, the first column gives the number of times the
maximum occurred within the 1st neighborhood, and the second column
gives the number of times the maximum occurred outside the 1st but
within the 2nd neighborhood. It is quite clear from the table that
the maximum did occur closest to the true locations in a overwhelmingly large majority of cases,
underscoring the effectiveness of the proposed method. These results
give us the required confidence in our estimation. See Figure \ref{FIG:cv_graph} for a 
visual idea about the locations of the maxima during cross-validation. The points 
which are chosen for the implementation of the cross-validation algorithm are 
indicated in red; green lines join them to the point where the corresponding 
maximum of the affinity was observed. 

\begin{table}[htbp]
\caption{Results of cross-validation.}
\label{table3}
\begin{center}
\begin{tabular}{ccc}\hline
 Model & 1st nbhood & 2nd nbhood\\ \hline\hline
 18sp3bar3 & 24 & 0\\ \hline
 25sp3bar3 & 24 & 0\\ \hline
 sp3bar3 & 24 & 0\\ \hline
 bar6 & 22 & 1\\ \hline
\end{tabular}
\end{center} 
\end{table}

In addition to performing cross-validation to check against internal
inconsistencies, we have successfully compared our results with those
reported by \citet{chak3} on the basis of their Bayesian method that
is independent of density estimation (see Section~\ref{SEC:confidence}).

\subsection{Direct Divergence Estimation}\label{SEC:direct_divergence}
\begin{figure}[htbp]
\includegraphics[scale=0.7]{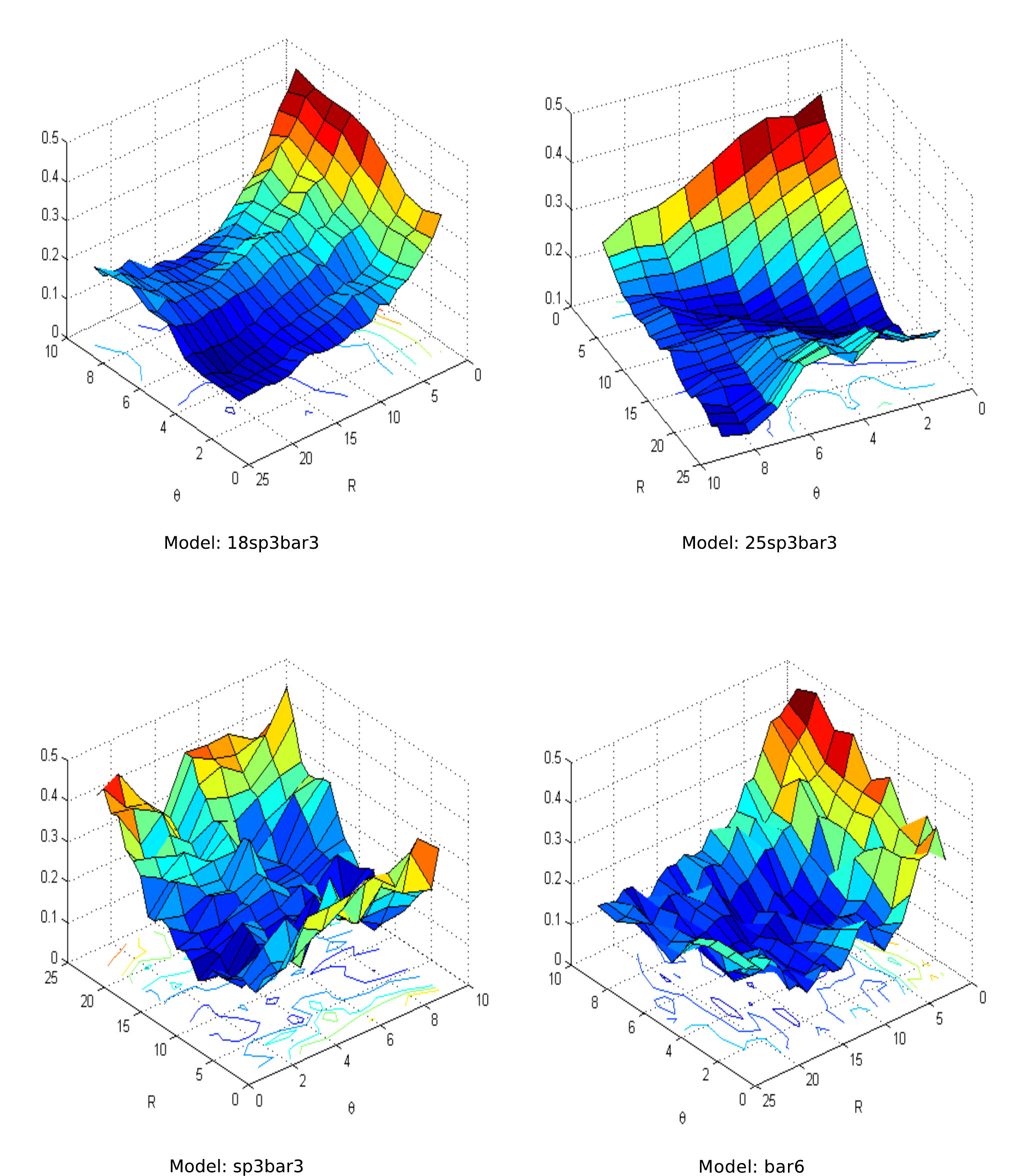}
\caption{Direct rPED surfaces.}
\label{FIG:rdiv_surfaces}
\end{figure}

On the suggestion of one of the reviewers, we explored some methods of construction of divergences avoiding density estimation. In particular we considered the construction of the rPE divergence (introduced in Section~\ref{SEC:rPE_div}) using the direct method of estimating the density ratio. The surfaces of the new divergence (Figure~\ref{FIG:rdiv_surfaces}) have reasonable similarity with the KLD surfaces (Figure~\ref{FIG:KLD_surfaces}), and general conclusions based on the new surfaces are largely compatible with our previous findings. Thus it appears that the methods that bypass the issue of density estimation can have some real utility in practice. We note, however, that at present theoretical consistency results about the direct density ratio method are limited in number as well as scope.

\section{Concluding Remarks}\label{SEC:conclusion}

\paragraph{}
In this paper we have developed a new method for estimating the
location of the Sun with respect to the center of the Milky
Way. Observed 2-dimensional velocity vectors of stars were used to
estimate the distribution of the observed stellar motion where the
location of the observer, i.e. the location of the Sun, with respect to
the center of the Galaxy is unknown. This distribution was compared to
distributions estimated using synthetic stellar velocities generated
at known locations in the Milky Way disk, where such synthetic data
were taken from the astronomical literature. The comparison was
performed by considering affinity measures based on the Hellinger
distance. In doing so we have made a direct determination of the
compatibility of the location from which the observed stellar
velocities were recorded, with these synthetic data sets. Our
procedure allows us to estimate the observer location directly as a
point on the (radial, angular) plane, rather than estimating the
components of the location vector individually. Indeed, the confidence
set of the estimated positions that we develop, based on the bootstrap
technique, is a set of locations on the $2$-dimensional plane rather
than a product of intervals. As a final test we run a consistency
check on the estimates through a cross-validation experiment which
indicates that the estimation procedure has some desirable continuity
properties. The method provides a new perspective on the problem under
consideration without contradicting the general belief about the
behavior of the astronomical models under study.

\section{Acknowledgments}\label{SEC:ackn}
The authors gratefully acknowledge the detailed comments of two anonymous referees, which led to a significantly improved version of the paper. The major part of the work was done when the first and the last named authors were Masters students at the Indian Statistical Institute, Kolkata.
\bibliographystyle{apalike}
\bibliography{milky_way}
\end{document}